\documentclass[usenatbib]{mn2e}
\input epsf

\newcommand{\rd}{{\rm d}}
\newcommand{\beq}{\begin{equation}}
\newcommand{\eeq}{\end{equation}}
\newcommand{\beqary}{\begin{eqnarray}}
\newcommand{\eeqary}{\end{eqnarray}}
\title[Serendipitous TNO occultation in X-rays]{Search for
serendipitous TNO occultation in X-rays} 
\author[Chang, Liu, and Chen]{Hsiang-Kuang Chang$^{1,2}$\thanks{E-mail:
hkchang@phys.nthu.edu.tw}, Chih-Yuan Liu$^{2,3}$, and Kuan-Ting Chen$^2$ 
\\
$^{1}$Institute of Astronomy, National Tsing Hua University, 
Hsinchu 30013, Taiwan\\ 
$^{2}$Department of Physics, National Tsing Hua University, 
Hsinchu 30013, Taiwan\\
$^{3}$LESIA, Paris Observatory, 92195 Meudon, France} 

\begin{document}

\date{Accepted November 19, 2012}

\pagerange{\pageref{firstpage}--\pageref{lastpage}} \pubyear{2012}

\maketitle

\label{firstpage}

\begin{abstract}
To study 
the population properties of small, remote objects 
beyond Neptune's orbit in the outer solar system, of kilometer size or smaller,
serendipitous occultation search is so far the only way.
For hectometer-sized Trans-Neptunian Objects (TNOs), optical shadows actually 
disappear because of diffraction.
Observations at shorter wave lengths are needed.
Here we report the effort of TNO occultation search in X-rays 
using RXTE/PCA data of Sco X-1 taken from June 2007 to October 2011.
No definite TNO occultation events were found in the 334 ks data. 
We investigate the detection efficiency dependence  
on the TNO size to better define the sensible size range of our approach 
and suggest upper limits to the TNO size distribution in the size range from
30 m to 300 m.  
A list of X-ray sources suitable for future larger facilities to observe
is proposed. 
\end{abstract}


%SSSSSSSSSSSSSSSSSSSSSSSSSSSSSSSSSSSSSSSSSSSSSSSSSSSSSSSSSS
\section{Introduction}
Serendipitous Trans-Neptunian Object (TNO) occultation search 
is so far the only way to explore properties of kilometer-sized TNOs or
smaller. 
Such search has been being conducted mainly
in optical bands
\citep{bianco10,bickerton08,roques03}
with few reported detections
\citep{roques06,schlichting09,schlichting12}.
In the X-ray band,
occultation by smaller objects may be 
detected because of a smaller Fresnel scale.
Following an earlier discovery of putative occultation events, 
allegedly caused by 100-meter size TNOs,
in the 1996-2002 X-ray data of Sco X-1 taken
by the instrument Proportional Counter Array (PCA)
on board Rossi X-ray Timing Explorer (RXTE)
\citep{chang06,chang07},
we conducted new RXTE/PCA observations of Sco X-1 from June 2007 to Oct 2011
to clarify possible instrumental-effect contamination in that discovery.

In this paper we report the final result of the effort in 
the search for serendipitous TNO occultation in X-rays with RXTE/PCA,
taking into account the issue of detection efficiency.
Earlier results based on data taken by Oct 2009 have been published
in \citet{liu08} and \citet{chang11}.
Currently such search is only feasible with RXTE/PCA obseravtion of Sco X-1,
because of the large effective area of PCA and because Sco X-1 is the brightest
X-ray source in the sky \citep{chang06,chang07}. Since RXTE
was decommissioned in January 2012, new efforts have to await possible
larger instruments in the future.

TNOs larger than deca-kilometer size can be detected directly. 
Results of those observations can be found in \citet{fuentes10} and 
\citet{fuentes11}
for works using HST, in \citet{fraser10} for Subaru, and in
\citet{petit11} for CFHT. The TNO size distribution, from thousand-kilometer
to decameter size and smaller, carries infromation of the collisional and
dynamical history of the early solar system. It is of essential importance
to our understanding of how the solar system formed
(see e.g. \citet{chiang07}, \citet{benavidez09}, \citet{fraser09} 
and \citet{kenyon12}).    

%SSSSSSSSSSSSSSSSSSSSSSSSSSSSSSSSSSSSSSSSSSSSSSSSSSSSSSSSS
\section{Summary of the RXTE/PCA 2007-2011 observations}
RXTE had observed Sco X-1 for many times, and yielded 
a large amount of data by
2007 (for details of RXTE instruments, see \citet{jahoda06}). 
Many millisecond-dip events were found in those data \citep{chang06}, 
but later it was pointed out that `Very Large Events' (VLEs)
 are likely the cause of those dips, 
rather than TNO occultations \citep{chang07,jones08}.
VLEs are events that deposit more than 100 keV into an individual anode in a
Proportional Counter Unit (PCU).
RXTE/PCA consists of 5 identical PCUs.
An instrument dead-time of 50 $\mu$s is set for each VLE 
during RXTE observations of Sco X-1. 
VLEs are produced by high-energy photons or particles. 
Whether they are really the cause of the millisecond dip events found earlier,
however, was not conclusive because VLEs were only
recorded in housekeeping data
with a coarse time resolution of 125 ms in the data taken by 2007.
In order to clarify the possible instrumental effect that may have 
caused the millisecond dip events found in \citet{chang06,chang07}, 
new observations with a newly designed data mode to record 
more detailed information of each individual VLE, 
such as the identification of the triggered  
PCU and the triggered anode and timing with
125 $\mu$s accuracy, were
conducted from June 2007 to October 2011, which produced in total
334 ks data good for the search for serendipitous TNO occultations.

As already reported in \citet{chang11}, 
if one distinguishes VLEs into different types according to
the number of anodes that were triggered, one can find that indeed
millisecond dip events are associated with VLEs of different types in
a systematic manner. 
The average count rate for all VLEs altogether in the 334-ks data is
$90.3\pm 18.8$ counts per second per PCU. 
The rate for each type of VLEs
is shown in Table \ref{vletype}. 
To see the association of dip events with VLEs, we divide dip events
into groups according to their associated VLE types, as shown in
Table \ref{dipgrp}.
For more detailed description of these classifications, as well as
the algorithm that we used to find and define the dip event,
see \citet{chang11}.
 
%TTTTTTTTTTTTTTTTTTTTTTTTTTTTTTT
\begin{table}
\begin{center}
\begin{tabular}{clc}
\hline
Type & number of triggered anodes & average count rate \\
\hline
A & no  & $1.73\pm 0.29$\\
B & all  & $34.5\pm 11.0$\\
C & more than one but not all & $44.7\pm 9.97$\\
D & only one & $9.37\pm 2.17$\\
\hline
\end{tabular}
\end{center}
\caption{VLE types and their average count rates (in units of counts
per second per PCU). The number of PCUs which are on during the observation
varies from time to time. The total data employed
in this analysis is 334 ksec, and is 1363 ksec-PCU when the number of PCU on
is taken into account. }
\label{vletype}
\end{table}
%TTTTTTTTTTTTTTTTTTTTTTTTTTTTTTT

%TTTTTTTTTTTTTTTTTTTTTTTTTTTTTTT
\begin{table}
\begin{center}
\begin{tabular}{llcc}
\hline
Group &  associated VLE type  & significant  & less significant  \\
 & & dips & dips \\
\hline
A & Type A & 45 & 180 \\
B & Type B, no Type A & 2 & 142 \\
C & Type C, no Type A, B & 3 & 35 \\
D & Type D only & 0 & 6 \\
E & no VLEs & 1 & 10 \\
\hline
\end{tabular}
\end{center}
\caption{Number of dips in different groups. `Significant' dips are those with
the number of counts smaller than $-6.5\sigma$ below the average 
in an 8-second running window, 
where $\sigma$ is the standard deviation of the counts in 
each bin in the running window. `Less significant' dips are those below
$-5.0\sigma$ but higher than $-6.5\sigma$.}
\label{dipgrp}
\end{table}
%TTTTTTTTTTTTTTTTTTTTTTTTTTTTTTT

From Table \ref{dipgrp} we can see that 45 of the 51 significant dips are of
Group A. Type-A VLE only has a count rate of 
$1.73\pm 0.29$ counts per second per PCU. 
Apparently Type-A VLEs are very strong events that cause these
dips, although exact details and why they are recorded with no anode triggered
are not yet understood.
The issue whether Type-B, -C, and -D VLEs can also 
produce millisecond dips was discussed in \citet{chang11} based on
the number of `less significant' dips.
In our search algorithm, we in fact rounded off the value of count deviations
to the first decimal, so all those dips with deviation between $-4.95\sigma$ 
and $-6.45\sigma$ were counted as `less significant' dips.
Since most of the dips are of 2-ms duration, we consider only the case of
2-ms bins 
(more precisely speaking, each 2-ms bin in fact is $8/4096$ s $= 1.95$ ms).
There are $1.71\times 10^8$ '2-ms' bins in total in the 334-ks data.
The expected number of  `less significant' dips due to
random fluctuation
is 63.4 if a Gaussian distribution is assumed.
The deviation distribution of the data is a dead-time-corrected
Poisson distribution, which is about a factor of $2\sim 3$ smaller than
a Gaussian at about $-5\sigma$; 
see the discussion and Figure 1 in \citet{chang11}.  
We therefore consider the total number of less significant
dips to be $63.4/2.5=25.4$.
Among the $1.71\times 10^8$ bins, $9.1\times 10^7$ bins are without the
occurrence of any VLEs.
The number of these 25.4 dips in Group E can be estimated to be
$25.4\times 9.1/17.1=13.5$.
That of the other groups can be estimated with
 the corresponding VLE count rates.
We therefore expect the numbers 
of `less significant' dips to be 0.23, 4.54, 5.88, 1.23, and 13.5
for
Group A to E respectively. 
From Table \ref{dipgrp} one can see
that only that of Group E is consistent with random fluctuation.
The numbers of `less significant' dips of other groups are clearly 
larger. The deviation is the largest for Group A, and gets smaller and smaller
from Group A to D. 
The above estimate is only a rough one because 
(1) we consider only the case of 2-ms bins, i.e. treating
 all the dips as of duration 2 ms, 
(2) in our definition of dip groups, Type B VLEs can also occur in Group A
dips, similarly for Type C and D VLEs, 
so the use of each VLE count rates for the estimation 
is not an accurate one, and
(3) the factor 2.5 to account for the difference from a Gaussian distribution
 is somewhat arbitrary.
Nonetheless, we think the above estimate is good enough to indicate
the VLE-association of those dips in Group A to D. 

Those significant dips are not due to random fluctuation.
The above discussion strongly suggests that those of Group A to D
are related to the occurence of VLEs and are therefore instrumental 
due to a process not yet fully understood.
The `significant' dip event in Group E, 
denoted as `Event E1' in \citet{chang11}, 
is the only non-random dip event that shows no indication of
any instrumental effect.
Its light curve was fit with shadow diffraction patterns in
\citet{chang11} and the result suggests that it is probably not due to
TNO occultation either. 

In this section we update the results presented in \citet{chang11}
and have the same conclusion: 
(1) The fact that most of the `significant' dip events are in Group A indicates
that Group-A dips are instrumental. 
(2) The numbers of `less significant' dip events of Group A, B, C and D
are larger than expected from random fluctuation. It strongly  
suggests that VLEs, even for Type D, which triggers only one anode, 
can result in millisecond
dip events. Only those dips without any association with VLEs can be
considered non-instrumental.

%SSSSSSSSSSSSSSSSSSSSSSSSSSSSSSSSSSSSSSSSSSSSSSSSSSSSSSSSS
\section{The detection efficiency}
In our earlier study to infer the TNO size distribution, 
a 100\% detection efficiency was assumed.
To better understand the detection efficieny of our approach,
we conducted simulations to investigate the dependence of
detection efficiency on the size and relative transverse speed of
the occulting body. The detection efficiency surely also 
depends on the observed count rate, which is related to 
the brightness of the occulted
background source and the instrument employed.
In the followng we consider two different count rates.
One is $8\times 10^4$ cps, which is close to what we usually have for
RXTE/PCA onservations of Sco X-1.
The other is $2\times 10^6$ cps, which is about 20 times more than RXTE/PCA
and could be achieved by future missions such as the proposed LOFT project
\citep{mignani11}.
The orbital speed of the Earth is about 30 km/s. That at 40 AU is 5 km/s.
The orbital speed of the RXTE spacecraft relative to the Earth is 
7.8 km/s. Sco X-1 is in a direction at about 
$5.5^\circ$ to the north of the ecliptic.
The relative transever (to the line of sight) speed of an occulting body
is usually largest when Sco X-1 is in opposition (around end May)  
and smallest when the Earth moves towards or away from the direction
of Sco X-1. The orientation of the motion of RXTE and of the occulting body 
will also affect that speed. We therefore consider three different speeds,
i.e. 5, 15 and 25 km/s, in the simulation respectively.

\begin{figure}
\epsfxsize=8.4cm
\epsffile{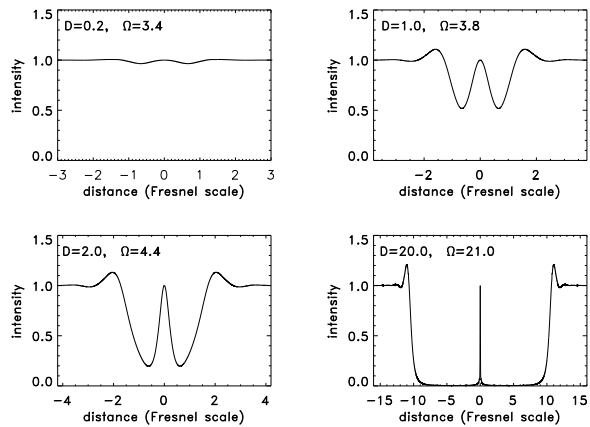}
\caption{Computed shadow light curves for occulting bodies of different
diameter $D$. All the 4 panels are for the case of a central-crossing event,
that is, for a zero impact parameter $\beta$. 
A typical RXTE/PCA observed Sco X-1 spectrum 
is used for the diffraction pattern computation. $\Omega$ is the shadow width
as defined in \citet{nihei07}. $D$ and $\Omega$ are both in units of a
Fresnel scale. 
}
\label{mdlc}
\end{figure}

For a given count rate and a given transverse speed, we would like to know the
chance of detecting an occultation event caused by an occulting body
of a certain size with our searching algorithm.
Most of the RXTE/PCA-observed photons from Sco X-1 are about 4 keV in energy
(0.3 nm in wavelength).
The Fresnel scale, which is defined as $\sqrt{\lambda d/2}$,
where $\lambda$ is the wavelength and $d$ the distance,
is 30 m for $\lambda=0.3$ nm and $d=40$ AU.
One expects to detect the shadow of an occulting body much larger 
than the Fresnel scale easily, while that of a body much smaller than 
the Fresnel scale will elude being detected due to diffraction. 
Another factor significantly affecting the observed shadow (light curve)
is the impact parameter $\beta$, usually defined to be the shortest distance of 
the passing path to the shadow center in units of the occulting body radius. 
A few shadow diffraction patterns, computed with the recipes in \citet{roques87}
and with a typical RXTE/PCA-observed Sco X-1 photon spectrum, are shown in
Figure~\ref{mdlc}. 
%FFFFFFFFFFFFFFFFFFFFFFFFFFFFFFFFFFFFFFFFFFFFFFFFFFFFFFF
%FFFFFFFFFFFFFFFFFFFFFFFFFFFFFFFFFFFFFFFFFFFFFFFFFFFFFFFF 
We note that with a point background source the diffraction patterns,
expressed in units of the Fresnel scale, 
are all the same for a fixed ratio of the occulting body size 
to the Fresnel scale. That is, there is a size-distance degeneracy.
Attempts to break this degeneracy are discussed in the next section.
We adopt the definition of the shadow boundary employed in 
\citet{nihei07}, in which the relation between
the shadow size $\Omega$ and the occulting body radius $\rho$ is
described as $\Omega=2(\sqrt{3}^{3/2}+\rho^{3/2})^{2/3}$, 
in which
both $\Omega$ and $\rho$ are in units of a Fresnel scale.  
Throughout this paper, when referring to the occulting body size we will
use the diameter $D$, which is obviously equal to $2\rho$.

In our simulation, with a given count rate $F$, a given transverse speed $v$,
and a given occulting body diameter $D$,
we randomly pick an impact parameter $\beta$ in the range 
from the shadow center ($\beta=0$) to its boundary 
($\beta=\frac{\Omega}{D}$) 
with a uniform probability and a random epoch 
for each individual implanted event
in a 100-s segment.
We first compute such a 100-s model light curve, then
bin the light curve into 0.25-ms bins, and then randomize the counts
in each bin to produce a simulated binned light curve.
This light curve is then processed with exactly the same algorithm that
we apply to real data.
We repeat the procedure $10^5$ times for each set of $F$, $v$ and $D$, and 
record the number of times of detection, which we use to represent the
detection efficiency.

%FFFFFFFFFFFFFFFFFFFFFFFFFFFFFFFFFFFFFFFFFFFFFFFFFFFFFFF
\begin{figure}
\epsfxsize=8.4cm
\epsffile{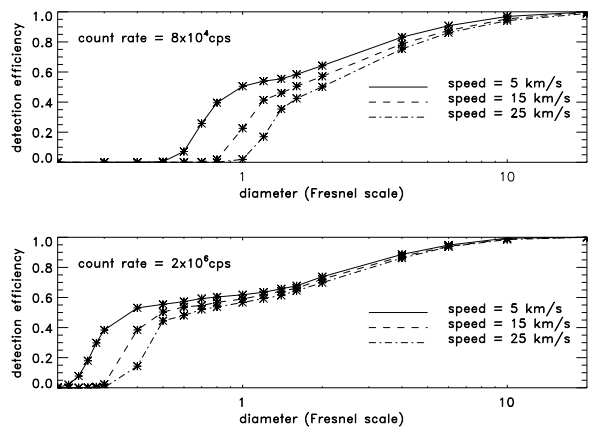}
\caption{The detection efficiency of our algorithm in the search for
 serendipitous TNO occultations as a function of the occulting body diameter. 
Three different relative transverse speeds are considered. 
We assume these occulting bodies are at 40 AU away, that is, one Fresnel scale
corresponds to 30 m.  
The upper panel is for the case of
the typical RXTE/PCA count rate, and the lower one is for a much higher
count rate, which may be achieved by future missions. 
}
\label{de}
\end{figure}
%FFFFFFFFFFFFFFFFFFFFFFFFFFFFFFFFFFFFFFFFFFFFFFFFFFFFFFFF 
The result of the detection efficiency determination is shown in
Figure~\ref{de}.
One can see that, with the current RXTE/PCA count rate, 
a detection efficiency larger than 50\% can be achieved for
occulting bodies larger than 2 Fresnel scale (diameter), considering the
case of relative transeverse speed being 25 km/s,
while for the high count rate case ($2\times 10^6$ cps), 
that occulting body size can be pushed down to about 0.6 Fresnel scale.  

%FFFFFFFFFFFFFFFFFFFFFFFFFFFFFFFFFFFFFFFFFFFFFFFFFFFFFFF
\begin{figure}
\epsfxsize=8.4cm
\epsffile{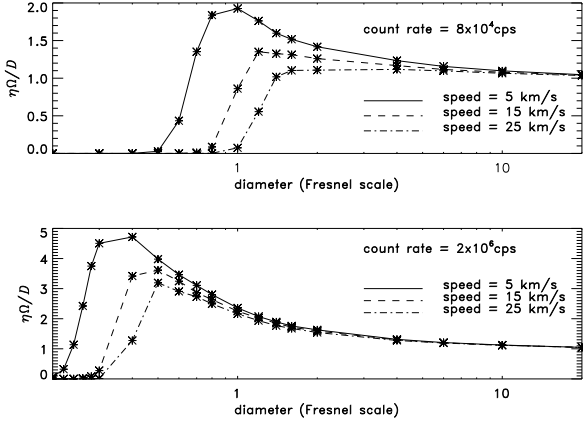}
\caption{The difference between the detection efficiency times shadow width 
($\eta\Omega$) and the occulting body diameter ($D$) 
expressed in the form of their
ratio as a function occulting body diameter.
For larger occulting bodies the fractional difference is small.
For smaller ones whose detection efficiency is not yet negligible 
it becomes large since the shadow width approaches a constant
value when the size decreases. 
}
\label{etomds}
\end{figure}
%FFFFFFFFFFFFFFFFFFFFFFFFFFFFFFFFFFFFFFFFFFFFFFFFFFFFFFFF 
%FFFFFFFFFFFFFFFFFFFFFFFFFFFFFFFFFFFFFFFFFFFFFFFFFFFFFFF
\begin{figure}
\epsfxsize=8.4cm
\epsffile{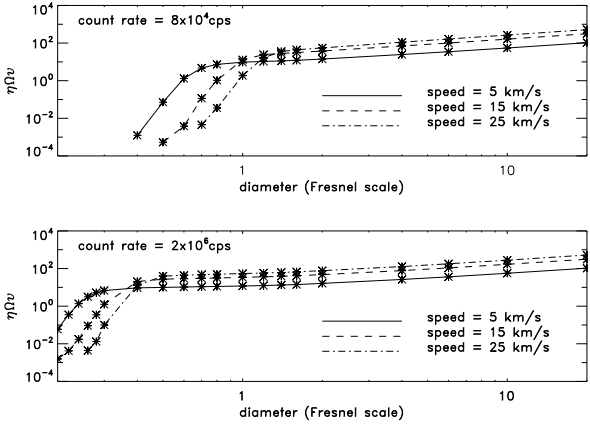}
\caption{The product $\eta\Omega v$, in units of the Fresnel scale times
km/s, as a function of 
the occulting body diameter.
This product is a relative contribution to the event rate estimate from TNOs
of different size and different transverse speeds. 
}
\label{etomv}
\end{figure}
%FFFFFFFFFFFFFFFFFFFFFFFFFFFFFFFFFFFFFFFFFFFFFFFFFFFFFFFF 
For a background point source, the event rate can be estimated as
(cf.\ Equation (8) in \citet{chang11})
\begin{equation}
\label{eventrate}
\frac{N}{T}=\frac{\int_{D_1}^{D_2}\left(\frac{{\rm d}n}{{\rm d}D}\right)
\eta\Omega v\,{\rm d}D}{d^2}\times(\frac{180}{\pi})^2 \,\,\,\, ,
\end{equation}
where $N$ is the number of detected events, 
 $T$ the total exposure time,
$\left(\frac{{\rm d}n}{{\rm d}D}\right)$ the differential size distribution
(in units of number per unit length per square degree),
$\eta$ the detection efficiency,
$\Omega$ the shadow size,
$v$ the typical sky-projection relative speed, 
 and $d$ the typical distance to the TNOs.
The product $\eta\Omega$ in the above equation
was approximated by the diameter in Chang et al. (2011)
The 
ratio of
them as a function of occulting body diameter $D$
is shown in Figure~\ref{etomds}.
The product $\eta\Omega v$ represents a certain relative detection probability.
It is plotted in Figure~\ref{etomv} as a function of $D$ for the three
different speeds.
We can see that
the product $\eta\Omega v$ becomes one order smaller when
the object diameter decreases from 20 Fresnel scale to 2 Fresnel scale.
For the current RXTE/PCA count rate,
it actually becomes relatively negligible for diameter smaller
than 1 Fresnel scale.
Nonetheless, $\frac{\rd n}{\rd D}$ as a function
of $D$ could be very steep, so the contribution   
to the event rate from objects of diameter between 1 and 2 Fresnel scale
could be considerable.
For the high count rate case, $\eta\Omega v$ becomes relatively
negligible when
the object diameter is smaller than 0.4 Fresnel scale.

We estimate the upper limit to the TNO size distribution 
at the level of setting $N=1$ in
Equation (\ref{eventrate}).
Although different models in the literature 
give different predictions for the
TNO size distribution of 
 sizes smaller than about 100 km, 
a common trait among most models is a wavy shape, 
caused by the competetion of
coagulation, erosion and shattering in different size ranges, 
down to a small size of about 0.1 km \citep{kenyon12}, 0.5 km \citep{fraser09},
or 1 km or so \citep{benavidez09}, below which the collisional equilibrium 
is achieved.  
In the collisional equilibrium, one expects to have
\citep{obrien03}
$\frac{\rd n}{\rd D}\sim D^{-q}$ and
\beq
q=\frac{7+p/3}{2+p/3}
\,\, ,
\eeq
where $p$ is the power index of the strength-size relation.
In the so-called `strength-scaled regime' (in contrast to the 
`gravity-scaled regime' for larger bodies), $p$ is negative. 
Most models give a $q$ in the range from 3.5 to 4.0 for
 the collisional equilibrium.
The distribution can be quite flat at the small-size end of the wavy-shape 
distribution. In \citet{kenyon12} $q$ is 
about 1.0 in the range from 0.1 km to 1 km.
If we describe the TNO size distribution in terms of
\beq
\frac{\rd n}{\rd \log D}
=\left(\frac{\rd n}{\rd \log D}\right)_{D=D_0}
\left(\frac{D}{D_0}\right)^{-q+1}
\,\,\, ,
\eeq
we can have, from Equation (\ref{eventrate}),
\beq
\left(\frac{\rd n}{\rd \log D}\right)_{D=D_0}
=\frac{1}{\int_{D_1}^{D_2}\left(\frac{D}{D_0}\right)^{-q+1}
\frac{\eta\Omega}{D}\rd D}\frac{N}{T}\frac{d^2}{v}\,(\frac{\pi}{180})^2\ln 10 
\,\, ,
\eeq
where $v$ is assumed to be a typical speed independent of $D$.

We consider two size ranges, that is, object diameter between 1.0 and 2.0 
Fresnel scale and between 2.0 and 10.0 Fresnel scale.
Assuming a typical distance $d=40$ AU and 
a typical relative sky-projection speed
$v=25$ km/s, 
with $T=334$ ks and setting one detection as the upper limit, 
we have in the size range from 30 m to 60 m 
(recall that one Fresnel scale is 30 m in the current case and
whenever we use 'size' we mean the diameter) 
$\left(\frac{{\rm d}n}{{\rm d}\log D}\right)_{D=30\,{\rm m}}
< 4.0\times 10^{11}\,\,\mbox{deg}^{-2}
$ for $q=4.0$ and 
$\left(\frac{{\rm d}n}{{\rm d}\log D}\right)_{D=30\,{\rm m}}
< 3.3\times 10^{11}\,\,\mbox{deg}^{-2}
$ for $q=3.5$.
In the size range from 60 m to 300 m we have 
$\left(\frac{{\rm d}n}{{\rm d}\log D}\right)_{D=60\,{\rm m}}
< 6.6\times 10^{10}\,\,\mbox{deg}^{-2}
$ for $q=4.0$,  
$\left(\frac{{\rm d}n}{{\rm d}\log D}\right)_{D=60\,{\rm m}}
< 5.7\times 10^{10}\,\,\mbox{deg}^{-2}
$ for $q=3.5$, and  
$\left(\frac{{\rm d}n}{{\rm d}\log D}\right)_{D=60\,{\rm m}}
< 1.2\times 10^{10}\,\,\mbox{deg}^{-2}
$ for $q=1.0$.  
Another factor of 1.14 should be applied to 
convert the estimate at the Sco X-1 latitude to the ecliptic;
see the discussion right after Equation (11) in Chang et al. (2011)
%FFFFFFFFFFFFFFFFFFFFFFFFFFFFFFFFFFFFFFFFFFFFFFFFFFFFF
\begin{figure}
\epsfxsize=8.4cm
\epsffile{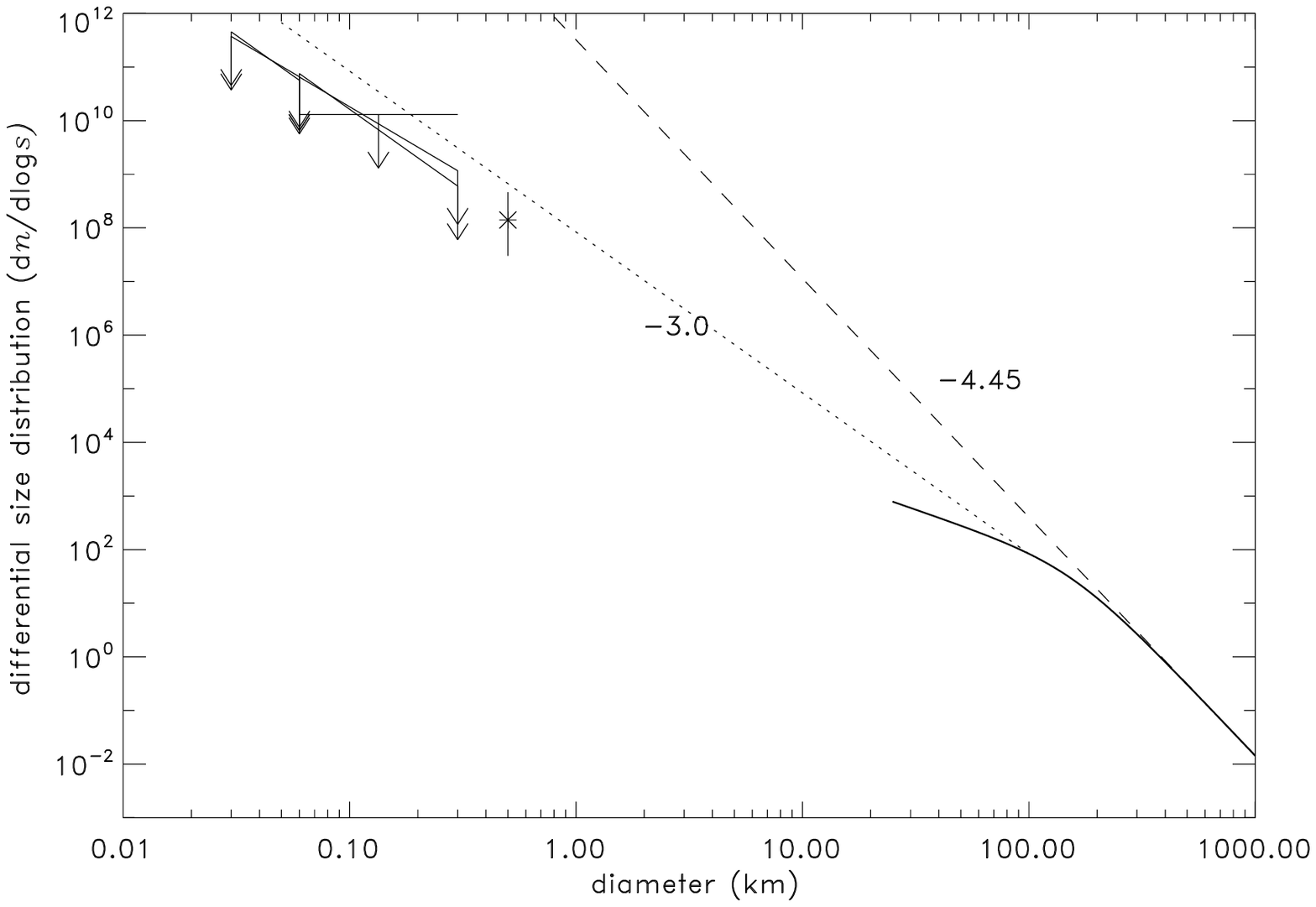}
\caption{The TNO size distribution. 
Plotted here is the differential sky surface density 
at the ecliptic per decade of size in units of number per square degree. 
The upper limits in the size range of 30-60 m and of 60-300 m, 
denoted with downward arrows, are derived from our non-detection 
in the 334-ks RXTE/PCA data of Sco X-1, 
and is set at the level of one detection in 334 ks. 
The cases of $q$ equal to 3.5 and 4.0 ($\frac{\rd n}{\rd\log D}\sim D^{-q+1}$)
are considered in the two size ranges, of which the derived upper limits
are very close to each other as plotted in this figure. 
The case of $q$ equal to 1.0 is also considered in the size range of 60-300 m. 
The downward arrows are plotted at the size-range boundary 
for $q$ equal to 3.5 and 4.0, and at the middle of the size range for 
$q$ equal to 1.0. 
The asterisk symbol at 0.5 km is based 
on the reported HST/FGS detection of an occultation event 
(\citet{schlichting09}; see also \citet{schlichting12}).
The solid curve is the double-power-law distribution of large TNOs
\citep{fuentes10}; see \citet{chang11} 
for the assumptions made to convert the luminosity function 
to the size distribution.
The dashed line is a direct extrapolation 
from the large size end of the double-power-law towards 
smaller size and the dotted line is a power law 
anchoring on the double-power-law at 90 km with a power index of -3.0. 
We note that the power indices in this figure, 
explicitly printed close to the corresponding power-law lines, 
are for the differential size distribution per decade of size, 
and are therefore the same as that of the cumulative size distribution. 
The double-power-law and the HST/FGS result in fact 
indicate a wavy shape for the TNO size distribution, 
similar to that of the main-belt asteroids. 
}
\label{sizedis}
\end{figure}
%FFFFFFFFFFFFFFFFFFFFFFFFFFFFFFFFFFFFFFFFFFFFFFFFFFFFFFFF
These upper limits are plotted in Figure~\ref{sizedis}.

%SSSSSSSSSSSSSSSSSSSSSSSSSSSSSSSSSSSSSSSSSSSSSSSSSSSSSSSSS
\section{Summary and discussion}
%TTTTTTTTTTTTTTTTTTTTTTTTTTTTTTTTTTTTTTTTTTTTTTTTTTTTTTTTTTTTT
\begin{table*}
\begin{center}
\begin{tabular}{lcccccl}
\hline
Sources & \multicolumn{3}{c}{Brightness (Crab)}   
&  $\lambda$ & $\beta$ & Other names   \\
\cline{2-4}
&  4U  & A1  &  ASMquick & & & \\ 
\hline
4U 1758-25 
& 1.21  & 0.56 &  0.58 
& 269.56 & -1.63 & 1H1758-250, GX5-1   \\

4U 1758-20
& 0.63  & 0.37 & 0.30 
& 269.66 & 2.91 &  1H1758-205, GX9+1, Sgr X-3 \\

4U 1617-15 
& 17.95  & 9.1 &  11.0        
& 245.14 & 5.73  & 1H1617-155, Sco X-1 \\ 

4U 1813-14
& 1.00  & 0.33 & 0.37 
& 273.24 & 9.35 & 1H1813-140, GX17+2, Ser X-2 \\ 

4U 1702-36 
& 0.79  & 0.41 & 0.44  
& 258.11 & -13.50 & 1H1702-363, GX 349+2, Sco X-2 \\ 
 
4U 1642-45 
& 0.48 & 0.18 & 0.25 
& 255.31 & -23.06 & 1H1642-455, GX340+0 \\

4U 1837+04
& 0.30  & 0.14 & 0.12 
& 280.60 & 28.09 & 1H1837+049, Ser X-1 \\ 

4U 2142+38
& 0.58  & 0.47 & 0.38  
& 345.98 & 47.96 & 1H2142+380, Cyg X-2 \\

4U 1956+35
& 1.24  & 0.29 & 0.42 
& 312.86 & 54.26 & 1H1956+350, Cyg X-1 \\
\hline
\end{tabular}
\end{center}
\caption{
Non-extended bright X-ray sources, which are potential 
background targets for serendipitous TNO occultation
search in X-rays. All the 9 listed sources are brighter 
than 0.1 Crab in all the 4th Uhuru (4U), HEAO 1 A-1 (A1),
and RXTE All Sky Monitor (ASMquick)
catalogues. The 4U catalog was compiled with observations 
conducted during 1972-1973, the A1 catalog was during 1977, 
and the ASMquick information is based on weekly average in August 2011. 
Those listed in the 2nd, 3rd and 4th columns are source brightness 
in the 4U, A1 and ASMquick catalogues respectively, all in units of a Crab. 
Their positions are in the 5th and 6th columns,
where $\lambda$ is the ecliptic longitude and
 $\beta$ the ecliptic latitude, both in degrees.
This list is sorted with the absolute value of the ecliptic latitude $\beta$.
}
\label{loftl}
\end{table*}
%TTTTTTTTTTTTTTTTTTTTTTTTTTTTTTTTTTTTTTTTTTTTTTTTTTTTTTTTTTTTT
%TTTTTTTTTTTTTTTTTTTTTTTTTTTTTTTTTTTTTTTTTTTT
\begin{table*}
\begin{center}
\begin{tabular}{lccccl}
\hline
Sources & \multicolumn{3}{c}{LOFT count rate} & Brightness & Type \\
\cline{2-4}
 & band (keV) & count rate (cps) & source fraction  & (Crab) & \\
\hline
4U 1758-25 &  
   1-2  & $4.8\times 10^3$ &  92.0\% 
& 1.02  & Z source, Cyg X-2 like \\
 & 2-4  & $1.3\times 10^5$ &  99.7\%  & & \\ 
 & 4-6  & $1.2\times 10^5$ &  99.7\%  & & \\ 
 & 6-10  & $8.3\times 10^4$ &   99.4\%  & & \\ 
 & 10-20  & $2.1\times 10^4$ &  95.4\%  & & \\ 
\hline
4U 1758-20 & 
   1-2  & $1.3\times 10^4$ &  97.0\% 
& 0.48 & Atoll source \\
 & 2-4  & $6.6\times 10^4$ &  99.5\%  & & \\
 & 4-6  & $5.1\times 10^4$ &  99.4\%  & & \\
 & 6-10  & $4.1\times 10^4$ &  98.8\%  & & \\
 & 10-20  & $1.2\times 10^4$ &  92\%  & & \\
\hline
4U 1617-15 & 
   1-2  & $1.8\times 10^6$ &  100\% 
& 12.6 & Z source (Sco X-1) \\
 & 2-4  & $2.5\times 10^6$ &  100\% &  & \\ 
 & 4-6  & $1.1\times 10^6$ &  100\%  &  & \\ 
 & 6-10  & $8.0\times 10^5$ &  99.9\%   & & \\ 
 & 10-20  & $2.4\times 10^5$ &  99.6\%   & & \\ 
\hline
4U 1813-14 & 
   1-2  & $6.9\times 10^3$ &  94.4\% 
& 0.50 & Z source, Sco X-1 like \\
 & 2-4  & $6.8\times 10^4$ &  99.5\%    & & \\ 
 & 4-6  & $5.6\times 10^4$ &  99.4\%    & & \\ 
 & 6-10  & $4.1\times 10^4$ &  98.8\%    & & \\ 
 & 10-20  & $1.4\times 10^4$ &  93.1\%    & & \\ 
\hline
4U 1702-36 & 
   1-2  & $1.8\times 10^4$  & 97.8\% 
& 0.66  & Z source, Sco X-1 like \\
 & 2-4  & $9.1\times 10^4$  & 99.6\%    & & \\ 
 & 4-6  & $7.1\times 10^4$  & 99.6\%     & & \\ 
 & 6-10  & $5.5\times 10^4$ &  99.1\%    & & \\ 
 & 10-20  & $1.7\times 10^4$ &   94.3\%    & & \\ 
\hline
4U 1642-45 & 
   1-2  & $1.3\times 10^3$ &  69.1\%
& 0.34 & Z source, Cyg X-2 like \\
 & 2-4  & $3.8\times 10^4$ & 99.1\%  & & \\ 
 & 4-6  & $4.4\times 10^4$ &  99.3\%  & & \\ 
 & 6-10  & $3.0\times 10^4$ &  98.3\%   & & \\ 
 & 10-20  & $7.3\times 10^3$ &  86.8\%   & & \\ 
\hline
4U 1837+04 & 
   1-2  & $1.8\times 10^4$ & 97.9\% 
& 0.21  & Atoll source \\
 & 2-4  & $3.3\times 10^4$ &  98.9\%  &  & \\ 
 & 4-6  & $2.0\times 10^4$ &  98.4\%  &  & \\ 
 & 6-10  & $1.7\times 10^4$ &  97.1\% &  & \\ 
 & 10-20  & $7.1\times 10^3$ &  86.4\%   & & \\ 
\hline
4U 2142+38 & 
   1-2  & $1.7\times 10^4$ &  97.8\% 
& 0.44 & Z source (Cyg X-2) \\
 & 2-4  & $7.6\times 10^4$ &  99.5\%  &  & \\ 
 & 4-6  & $4.6\times 10^4$ &  99.3\%  &  & \\ 
 & 6-10  & $3.0\times 10^4$ &  98.4\%  &  & \\ 
 & 10-20  & $8.6\times 10^3$ &  88.8\%  &  & \\ 
\hline
4U 1956+35 & 
   1-2  & $4.6\times 10^4$ &  99.1\% 
& 0.5 & black hole candidate, HMXB \\
 & 2-4  & $7.3\times 10^4$ &  99.5\% & & \\ 
 & 4-6  & $4.5\times 10^4$ &  99.3\% &  & \\ 
 & 6-10  & $4.4\times 10^4$ &  98.9\%   & & \\ 
 & 10-20  & $3.0\times 10^4$ &  96.8\%   & & \\ 
\hline
\end{tabular}
\end{center}
\caption{
Estimated LOFT count rates for the 9 sources listed in Table 1.
These count rates are estimated with LOFT response files and are
based on spectral models reported in
the literature 
(4U 1758-25: \citet{jackson09}; 
4U 1758-20: \citet{iaria05};
4U 1617-15: \citet{bradshaw03};
4U 2142+38: \citet{bachurch10};
4U 1956+35: \citet{nowak11};
the other 4 sources: \citet{cackett09}). 
Those listed count rates are the total count
rate (including background). 
The brightness, in units of a Crab, in the 5th column is based
on the assumed spectral model and is for the 2-10 keV   band,
in which a Crab is $2.4\times 10^{-8}$ erg cm$^{-2}$ sec$^{-1}$. 
}   
\label{loftr}
\end{table*}
%TTTTTTTTTTTTTTTTTTTTTTTTTTTTTTTTTTTTTTTTTTTTTTTTTTTTTTTTTTTTT
 
We conclude the current 
effort of serendipitous TNO occultation search 
in X-rays with upper limits to the TNO size distribution
in the size range from 30 m to 300 m. 
In RXTE/PCA observations of Sco X-1 from June 2007 to October 2011,
only one non-instrumental dip event was found in the 334-ks data.
Due to the distance-size degeneracy in the occultation light curve,
one cannot easily determine the occulting body size or distance.
One way to go around is to fit the light curve 
with different diffraction patterns
and find the best-fit result of the relative transverse speed,
which can be obtained only in units of Fresnel scale per unit time. 
Then with some orbital assumptions, this speed can be compared to
the real relative transverse speed of a body at a certain distance
to find the match \citep{chang11}.
As discussed in \citet{chang11},
based on the fitting result of the aforementioned approach, 
that non-instrumental dip event might be due to a TNO of 150-m size, 
but with a rare retrograde orbit,
or, it might be due to an MBA of 40-m size, but the implied size distribution
is incredibly steep, or, it might be due to a very nearby object 
of meter size moving at a relative speed of a few kilometers per second.
The main reason of why there exists such a huge uncertainty in the result is
the small number of photons detected at millisecond time scale.

Another possible way to break the distance-size degeneracy is to
relax the assumption of the background star being a point source.
The size of the X-ray emitting region in Sco X-1 is still under debate,
ranging from about 50 km to 50,000 km, depending on different models.
Sco X-1 is  at a distance about 2.8 kpc. 
The correspondig angular size of its X-ray emitting region
is 0.1 Fresnel scale at 40 AU considering the 50,000 km case.  
It is a good approximation to assume a point background source in such a case.
On the other hand, if we are confident on the knowledge of the emitting region
size and our instrument is large enough to provide good photon statistics,
with a given angular size of the background source,
one can detect the difference among diffraction shadow patterns 
caused by occulting bodies
of the same size in units of Fresnel scale at different distances,  
since the projection size of the background source at different distances
are different in units of Fresnel scale. 
For Sco X-1, if the occulting body is in the inner Oort Cloud, say, at 4000 AU,
its X-ray emitting region will appear to be 1-Fresnel scale large at that distance for
the 50,000-km case. 
From detailed study of the occultation light curve, one may be able 
to distinguish the projection size of the background source in units of Fresnel scale
and therefore to determine the distance. 
Again, this requires a larger instrument to provide good enough photon 
statistics.

After 16 years of fruitful service, RXTE was decommissioned in January 2012.
Future projects of building larger instruments have been proposed, 
such as Athena \citep{barcons12}, AXTAR \citep{ray11}, 
and LOFT \citep{mignani11}. In particular, LOFT will be 20 times larger
than RXTE/PCA in terms of its photon collecting power. 
In Table \ref{loftl} and \ref{loftr} we list 9 brightest non-extended 
X-ray sources suitable
for serendipitous TNO occultation search in the LOFT era.  
X-ray sources tend to be variable. Those 9 sources are all brighter than 
0.1 Crab in all the 4th Uhuru, HEAO 1 A1 and RXTE All Sky Monitor catalogues.
See table captions for more details. Those 9 sources are located at 
different ecliptic latitude, providing the possiblity to study latitude 
dependence of TNO distribution in the decameter to hectometer size range.
Their estimated LOFT count rates are also high enough to allow
investigation in two or three energy bands, which can help to
identify  diffraction features and to determine whether
 the discovered dip events are occultation in nature. 
One of the 9 sources is a High-Mass X-rar Binary (HMXB) containing a
black hole candidate. The other 8 are all Low-Mass X-ray Binaries (LMXBs),
two of which are the so-called `Atoll sources' and the other 6 are 
`Z sources'. They are all neutron-star systems showing various rich
variation in their timing and spectral properties. 
Long-term monitoring of these 9 sources is by itself very much rewarding for
the study of accretion physics around black holes and neutron stars, and
is at the same time useful for pinning down the TNO size distribution
in the decameter and hectometer size range.

%SSSSSSSSSSSSSSSSSSSSSSSSSSSSSS
\section*{Acknowledgments}
We thank Ed Morgan, who designed the new RXTE/PCA data mode 
and coordinated the new RXTE observations of Sco X-1. 
%This research has also made use of data obtained through 
%the High Energy Astrophysics Science Archive Research Center Online Service, 
%provided by the NASA/Goddard Space Flight Center. 
This work was supported by the National Science Council of 
the Republic of China under grants NSC 99-2112-M-007 -017 -MY3
and 101-2923-M-007 -001 -MY3.

\label{lastpage}
\end{document}